\documentclass[12pt,a4paper]{article}
\usepackage{jheppub}
\title{Observational constraints on massive gravity}

\author{Yungui Gong}
\affiliation{MOE Key Laboratory of Fundamental Quantities Measurement, School of Physics, Huazhong University of Science and Technology, Wuhan 430074, China}
\affiliation{Institute of Theoretical Physics, Chinese Academy of Sciences, Beijing 100190, China}
\emailAdd{yggong@mail.hust.edu.cn}

\abstract{
The ghost free massive gravity modified Friedmann equations at cosmic scale and provided an explanation
of cosmic acceleration without dark energy. We analyzed the cosmological solutions of the massive gravity
in detail and confronted the cosmological model with current observational data. We found that the model parameters $\alpha_3$
and $\alpha_4$ which are the coefficients of the third and fourth order nonlinear interactions cannot be
constrained by current data at the background level. The mass of graviton is found to be the order
of current Hubble constant if $\alpha_3=\alpha_4=0$, and the mass of graviton can be as small as possible
in the most general case.}


\arxivnumber{arXiv: 1210.xxxx}

\begin{document}

\maketitle

\section{Introduction}
A lot of efforts have been made to understand the accelerating expansion of the universe discovered by
the observations of Type Ia supernovae (SNe Ia) in 1998 \citep{acc1,acc2}. Although the economic explanation of the acceleration is a cosmological
constant which is consistent with all observations, the smallness of the cosmological constant and other problems such as the coincidence problem motivated the
modification of the theory of gravity. One way of modifying gravity is to add a small mass to graviton. In fact, Fierz and Pauli
made the first attempt to consider a theory of gravity with massive graviton \cite{pauli}. They add a quadratic mass term $m^2(h_{\mu\nu}h^{\mu\nu}-h^2)$
for linear gravitational perturbations $h_{\mu\nu}$ to the action, which breaks the gauge invariance of general relativity.
However, the linear theory with the Fierz-Pauli mass does not recover general relativity in the massless limit $m\rightarrow 0$, which leads to the
contradiction with solar system tests due to the vDVZ discontinuity  \citep{vdvz1,vdvz2}.
The discontinuity can be overcome by introducing nonlinear interactions with the help of Vainshtein mechanism \citep{vainshtein}.
Along this line, Dvali, Gabadadze and Porrati proposed a model of massive gravity in the context of extra dimensions which modifies general relativity
at the cosmological scale and admits a self-accelerating solution with dust matter only \cite{dgp}.

On the other hand, in the language of St\"{u}ckelberg fields, the nonlinear terms usually contain more than two time derivatives which
present the Bouldware-Deser (BD) ghost \citep{bdghost}. To remove the ghost, nonlinear interactions with higher derivatives are added order
by order in perturbation theory so that they re-sum to be a total derivative.
Recently, de Rham, Gabadadze and Tolley (dGRT)
successfully constructed a nonlinear theory of massive gravity \citep{massgrav} that is free from BD ghost \citep{hassan}.
Cosmological solutions with self acceleration for massive gravity were then sought by several groups \citep{amico,gong12,gumruk11,gratia12,koyama1,koyama2,kobayashi12,mukohyama,tolley,langlois12}.
Gumrukcuoglu, Lin, and Mukohyamafound found that a de-Sitter solution with an effective cosmological constant
proportional to the mass of graviton exists for a spatially open universe in the dGRT model of massive gravity \cite{gumruk11}.
The same solution was then found for spatially flat universe in \cite{gratia12}. When the parameters in the dGRT theory take
some particular values, Kobayashi, Siino, Yamaguchi and Yoshida found that the solution also existed for a universe
with arbitrary spatial curvature \cite{kobayashi12}. The same solution was obtained by different group with different method for some particular case, and
they all took the reference metric to be Minkowski.  Langlois and Naruko took a different approach by assuming the reference metric to be de-Sitter
and found more general cosmological solutions
in addition to the cosmological constant solution \cite{langlois12}. These new cosmological solutions
opened another door to the understanding of cosmic acceleration.
Cosmological solutions for the ghost-free bi-gravity were also found and confronted with observational data \citep{volkov11,volkov12,cardonrp,lambiase,vonStrauss:2011mq}.

In this paper, we focus on the cosmological solutions found in \cite{langlois12} for dGRT massive gravity \citep{massgrav}.
The Friedmann equations are modified so that it is possible to explain the cosmic acceleration.
We apply the SNLS3 SNe Ia data \citep{snls3}, the baryon acoustic oscillation (BAO) data \citep{wigglez} and the 7-year Wilson Microwave Anisotropy Probe (WMAP7) data \citep{wmap7}
to constrain the parameters in dGRT massive gravity.

\section{massive gravity}

In this paper, we study the ghost free theory of massive gravity proposed by \cite{massgrav},
\begin{equation}
\label{action}
S=\frac{M_{pl}^2}{2}\int d^4x\sqrt{-g}(R+m_g^2\mathcal{U})+S_m,
\end{equation}
where $m_g$ is the mass of graviton, the nonlinear higher derivative terms for the massive graviton is
\begin{gather}
\label{massv}
\mathcal{U}=\mathcal{U}_2+\alpha_3\mathcal{U}_3+\alpha_4\mathcal{U}_4,\\
\label{massv2}
\mathcal{U}_2=[\mathcal{K}]^2-[\mathcal{K}^2],\\
\label{massv3}
\mathcal{U}_3=[\mathcal{K}]^3-3[\mathcal{K}][\mathcal{K}^2]+2[\mathcal{K}^3],\\
\label{massv4}
\mathcal{U}_4=[\mathcal{K}]^4-6[\mathcal{K}]^2[\mathcal{K}^2]+8[\mathcal{K}^3][\mathcal{K}]-6[\mathcal{K}^4],
\end{gather}
and
\begin{equation}
\label{tensor1}
\mathcal{K}^\mu_\nu=\delta^\mu_\nu-(\sqrt{\Sigma})^\mu_\nu,
\end{equation}
The tensor $\Sigma_{\mu\nu}$ is defined by four St\"{u}ckelberg fields $\phi^a$ as
\begin{equation}
\label{stuckphi}
\Sigma_{\mu\nu}=\partial_\mu\phi^a\partial_\nu\phi^b\eta_{ab}.
\end{equation}
The reference metric $\eta_{ab}$ is arbitrary and it is usually taken to be Minkowski.
For an open universe, Gumrukcuoglu, Lin, and Mukohyamafound found the first cosmological solution with
an effective cosmological constant proportional to the mass of graviton \cite{gumruk11},
\begin{equation}
\label{efflambda}
\Lambda_{eff}=-m_g^2(X_\pm-1)[(1+3\alpha_3)X_\pm-3(1+\alpha_3)],
\end{equation}
where
\begin{equation}
\label{twosol1}
X_{\pm}=\frac{1+6\alpha_3+12\alpha_4\pm \sqrt{1+3\alpha_3+9\alpha_3^2-12\alpha_4}}{3(\alpha_3+4\alpha_4)}.
\end{equation}
It is obvious that $\alpha_3+4\alpha_4\neq 0$ for this solution, and two branches exist.
The same solution was then found in \cite{gratia12} for a flat universe.
It is natural to think that this solution should exist for a closed universe. If the parameters $\alpha_3$ and $\alpha_4$ take the particular value
\begin{equation}
\label{1parm}
\alpha_3=\frac{1}{3}(\alpha-1),\quad \alpha_4=\frac{1}{12}(\alpha^2-\alpha+1),
\end{equation}
the same cosmological constant solution
with $\Lambda_{eff}=m^2_g/\alpha$  independent of the curvature of the universe was found in \cite{kobayashi12}.
All these results are based on the assumption that the reference metric is Minkowski, and the method
of obtaining the solution cannot be generalized to the other case.

Langlois and Naruko took a different approach and assumed de Sitter metric for the reference metric \cite{langlois12},
\begin{equation}
\label{refmetric}
\eta_{ab}d\phi^a d\phi^b=-dT^2+b_k^2(T)\gamma_{ij}dX^idX^j,
\end{equation}
where the St\"{u}ckelberg fields are assumed to be $\phi^0=T=f(t)$, $\phi^i=X^i=x^i$,
so that the tensor $\Sigma_{\mu\nu}$ takes the homogeneous and isotropic form,
\begin{equation}
\label{stuckphi1}
\Sigma_{\mu\nu}={\rm Diag}\{-\dot{f}^2,\ b^2_k[f(t)]\gamma_{ij}\},
\end{equation}
and the functions $b_k(T)$ ($k=0$, $\pm 1$) are
$$b_0(T)=e^{H_c T},\quad b_{-1}(T)=H_c^{-1}\sinh(H_cT),\quad b_1(T)=H_c^{-1}\cosh(H_cT).$$

Varying the action (\ref{action}) with respect to the lapse function $N(t)$ and scale factor $a(t)$, we obtain Friedmann equations
\begin{gather}
\label{frweq1}
H^2+\frac{k}{a^2}=\frac{1}{3 M_{pl}^2}(\rho_m+\rho_g),\\
\label{frweq2}
2\dot H+3H^2+\frac{k}{a^2}=-\frac{1}{M_{pl}^2}(p_m+p_g),
\end{gather}
where the effective energy density $\rho_g$ and pressure $p_g$ for the massive graviton are,
\begin{gather}
\label{effrho2}
\begin{split}
\rho_g=\frac{m_g^2M^2_{pl}}{a^3}(b_k[f]-a)\{6(1+2\alpha_3+2\alpha_4)a^2-(3+15\alpha_3+24\alpha_4)a b_k[f]\\
+3(\alpha_3+4\alpha_4)b_k[f]^2\},
\end{split}\\
\label{effp2}
\begin{split}
p_g=&\frac{m_g^2M^2_{pl}}{a^3}\{[6+12\alpha_3+12\alpha_4-(3+9\alpha_3+12\alpha_4){\dot f}]a^2\\
&-2[3+9\alpha_3+12\alpha_4-(1+6\alpha_3+12\alpha_4){\dot f}]a b_k[f]\\
&+[1+6\alpha_3+12\alpha_4-3(\alpha_3+4\alpha_4){\dot f}]b_k^2[f]\}.
\end{split}
\end{gather}
Varying the action (\ref{action}) with respect to the function $f(t)$, we obtain three branches of
cosmological solutions \citep{langlois12}, the first two solutions $b_k[f(t)]=X_{\pm}\, a(t)$
correspond to the effective cosmological constant
and are independent of the explicit form of $b_k(f)$ as long as
the function $b_k(f)$ is invertible.\footnote{The effective cosmological constant (\ref{efflambda}) is obtained by substituting the solution
$b_k[f(t)]=X_{\pm}\, a(t)$ into the energy density of massive graviton (\ref{effrho2})}
The third solution is \citep{langlois12,tolley}
\begin{equation}
\label{masssol1}
\frac{db_k[f]}{df}=\frac{\dot a}{N}.
\end{equation}
For the flat case, $k=0$, substituting the de Sitter function $b_0[f(t)]=e^{H_c f(t)}$
into equations (\ref{masssol1}) and (\ref{effrho2}),
we obtain the effective energy density
and pressure for the massive graviton,
\begin{gather}
\label{effrho1}
\begin{split}
\rho_g=m_g^2M_{pl}^2\left[-6(1+2\alpha_3+2\alpha_4)+9(1+3\alpha_3+4\alpha_4)\frac{H}{H_c}\right.\\
\left.-3(1+6\alpha_3+12\alpha_4)\frac{H^2}{H_c^2}+3(\alpha_3+4\alpha_4)\frac{H^3}{H_c^3}\right],
\end{split}\\
\label{effp1}
\begin{split}
p_g=-\rho_g+m_g^2M_{pl}^2\frac{\dot H}{H^2}\frac{H}{H_c}\left[-3(1+3\alpha_3+4\alpha_4)+2(1+6\alpha_3+12\alpha_4)\frac{H}{H_c}\right.\\
\left.-3(\alpha_3+4\alpha_4)\frac{H^2}{H^2_c}\right].\\
\end{split}
\end{gather}
Note that when $H(z)=H_c$, $\rho_g=0$ and the contribution to the energy density from massive gravity is zero at this moment.
For the flat case, substituting equations (\ref{effrho1}) and (\ref{effp1}) into Friedmann equations (\ref{frweq1}) and (\ref{frweq2}), we get
\begin{equation}
\label{frweq4}
\begin{split}
\frac{m_g^2}{H_0^2}\frac{H(z)}{H_c}\left[-(\alpha_3+4\alpha_4)\frac{H^2(z)}{H_c^2}
+(1+6\alpha_3+12\alpha_4)\frac{H(z)}{H_c}-3(1+3\alpha_3+4\alpha_4)\right]\\
=-E^2(z)+\Omega_m(1+z)^{3(1+w_m)}-2\frac{m_g^2}{H_0^2}(1+2\alpha_3+2\alpha_4).
\end{split}
\end{equation}
\begin{equation}
\label{acceq4}
\begin{split}
\frac{\dot H}{H^2}\left\{-2E^2(z)+\frac{m_g^2}{H_c^2}E(z)\left[3(1+3\alpha_3+4\alpha_4)\frac{H_c}{H_0}-2(1+6\alpha_3+12\alpha_4)E(z)\right.\right.\\
\left.\left.+3(\alpha_3+4\alpha_4)\frac{H_0}{H_c}E^2(z)\right]\right\}=3\Omega_{m}(1+w_m)(1+z)^{3(1+w_m)},
\end{split}
\end{equation}
where $E(z)=H(z)/H_0$. The effective equation of state $w_g=p_g/\rho_g$ for the massive graviton is
\begin{equation}
\label{weff4}
w_g=-1-\frac{2E^2(z)\dot{H}/H^2+3\Omega_m(1+w_m)(1+z)^{3(1+w_m)}}{3[E^2(z)-\Omega_m(1+z)^{3(1+w_m)}]}.
\end{equation}
Without loss of generality, we assume that $m_g^2=-\beta_1 H_0^2$, and $H_c=\beta_2 H_0$.
Note that the mass appears in the action as a potential term, so the sign of $m_g^2$ can be absorbed
into the sign convention of the potential.
At the present time $z=0$, $E(0)=1$, equation (\ref{frweq4}) gives
\begin{equation}
\label{b1eq1}
\begin{split}
\beta_1=\frac{(1-\Omega_m)\beta_2^{3}}{2(1+2\alpha_3+2\alpha_4)\beta_2^{3}-3(1+3\alpha_3+4\alpha_4)\beta_2^{2}
+(1+6\alpha_3+12\alpha_4)\beta_2-(\alpha_3+4\alpha_4)}.
\end{split}
\end{equation}
If $\alpha_3+4\alpha_4=0$, the cubic term of Hubble parameter in equation (\ref{frweq4})
is absent and the cosmological evolution becomes simpler. Therefore we consider this special case first.
For the special case $4\alpha_4=-\alpha_3$, Friedmann equation (\ref{frweq4}) becomes
\begin{equation}
\label{frweq7}
\begin{split}
\left[1-\frac{\beta_1}{\beta_2^2}(1+3\alpha_3)\right]E^2(z)+3(1+2\alpha_3)\frac{\beta_1}{\beta_2}E(z)
=\Omega_m(1+z)^{3(1+w_m)}+\beta_1(2+3\alpha_3),
\end{split}
\end{equation}
with
\begin{equation}
\label{b1eqcase2}
\beta_1=\frac{(1-\Omega_m)\beta_2^2}{[(2+3\alpha_3)\beta_2-1-3\alpha_3](\beta_2-1)}.
\end{equation}
When $\beta_2\gg 1$, $\beta_1\approx (1-\Omega_m)/(2+3\alpha_3)$ and the standard $\Lambda$CDM model with cosmological constant $\Omega_\Lambda=1-\Omega_m$ is recovered.
Note that when $\beta_2\gg 1$, the model just weakly depends on the parameter $\alpha_3$ and $m^2_g\approx 0$ when $\alpha_3\gg 1$.
The deceleration parameter in this model is
\begin{equation}
\label{qzeq2}
q(z)=-1+\frac{3\Omega_m(1+w_m)(1+z)^{3(1+w_m)}E^{-2}(z)}{2-2(1+3\alpha_3)\frac{\beta_1}{\beta_2^2}+3(1+2\alpha_3)\frac{\beta_1}{\beta_2}E^{-1}(z)}.
\end{equation}
In this case, we have three model parameters $\Omega_m$, $\beta_2$ and $\alpha_3$.
Apparently, the coefficient of $H^2(z)$ should be positive, so the model parameters must satisfy the following condition
\begin{equation}
\label{2parcond1}
\frac{(1-\Omega_m)(1+3\alpha_3)}{[(2+3\alpha_3)\beta_2-1-3\alpha_3](\beta_2-1)}<1.
\end{equation}
At early times, $E(z)\gg 1$, the square term $E^2(z)$ dominates the left hand side of equation (\ref{frweq7}),
the standard cosmology with an effective cosmological constant is recovered and the effective matter density is $\Omega_m/[1-\beta_1(1+3\alpha_3)/\beta_2^2]$
instead of $\Omega_m$. To guarantee that equation (\ref{frweq7}) always has solutions, we require that
\begin{equation}
\label{2parcond2}
\Delta=9(1+2\alpha_3)^2\frac{\beta_1^2}{\beta_2^2}+4\left[1-\frac{\beta_1}{\beta_2^2}(1+3\alpha_3)\right]C(z)>0,
\end{equation}
where $C(z)=\Omega_m(1+z)^{3(1+w_m)}+\beta_1(2+3\alpha_3)$. Explicitly, the dimensionless Hubble parameter $E(z)$ is
\begin{equation}
\label{frwsol2.0}
E(z)=\frac{-3(1+2\alpha_3)\beta_1/\beta_2+\sqrt{\Delta}}{2[1-(1+3\alpha_3)\beta_1/\beta_2^2]}.
\end{equation}
Finally, to enure that $E(0)=1$, we require
\begin{equation}
\label{2parcond3}
-3(1+2\alpha_3)\frac{\beta_1}{\beta_2}<2-2(1+3\alpha_3)\frac{\beta_1}{\beta_2^2}.
\end{equation}

To better understand the dynamics, we analyze the simplest case
$\alpha_3=\alpha_4=0$ in more details, in which we have only two model parameters $\Omega_m$ and $\beta_2$. The Hubble parameter evolves as
\begin{equation}
\label{frwsol2.1}
E(z)=\frac{-3\frac{\beta_1}{\beta_2}+\sqrt{9\frac{\beta_1^2}{\beta_2^2}+4(1-\frac{\beta_1}{\beta_2^2})[\Omega_m(1+z)^{3(1+w_m)}+2\beta_1]}}{2(1-\beta_1/\beta_2^2)}.
\end{equation}
The condition (\ref{2parcond1}) is reduced to
\begin{equation}
\label{1parcond1}
\frac{(1-\Omega_m)}{(2\beta_2-1)(\beta_2-1)}<1.
\end{equation}
(1) If $1/2<\beta_2<1$, then $\beta_1<0$, $m_g^2>0$ and the above condition (\ref{1parcond1}) is automatically satisfied.
Combining this result with condition (\ref{2parcond2}), we get
$1/2<\beta_2<(3-\sqrt{\Omega_m})/4$ or $(3+\sqrt{\Omega_m})/4<\beta_2<1$ and $m_g^2>2H_0^2$. Since equation (\ref{frweq7}) is a quadratic equation,
there are two solutions, and we need to take the solution which $E(z)$ increases as the redshift $z$ increases and we require that
$E(z=0)=1$. To satisfy these requirements, the parameter $\beta_2$ must be in the region  $1/2<\beta_2<(3-\sqrt{\Omega_m})/4$.
(2) If $\beta_2>1$, then $\beta_1>0$ and $m_g^2<0$. The conditions (\ref{1parcond1})
and (\ref{2parcond2}) require that $\beta_2>(3+\sqrt{9-8\Omega_m})/4$. When $\beta_2\gg 1$, $\beta_1\approx (1-\Omega_m)/2$,
$m_g^2\approx -(1-\Omega_m)H_0^2/2$ and the standard
$\Lambda$CDM model is recovered. (3) If $0<\beta_2<1/2$, the conditions (\ref{1parcond1})
and (\ref{2parcond2}) require that $0<\beta_2<(3-\sqrt{9-8\Omega_m})/4$.
When $\beta_2\ll 1$, $\beta_1\approx (1-\Omega_m)\beta_2^2\ll 1$
and $m_g^2\approx 0$. However, the standard
$\Lambda$CDM model can not be recovered at early times. Therefore, we don't consider this case.

Now we consider the general case with $\alpha_3+4\alpha_4\neq 0$.
Since $\rho_g=0$ when $H(z)=H_c$, so if $H_c=H_0$, we find that $\Omega_m=1$ which
is inconsistent with current observations, so $H_c\neq H_0$. If $H_c<H_0$,
then in the past $z\gg 1$, the cubic term $H^3(z)$ dominates over the quadratic $H^2(z)$ and the linear $H(z)$ terms,
so we cannot recover the standard cosmology $H^2\sim \rho$ unless we fine tune the value
of $m_g^2/H_0^2$ to be very small. Therefore, we require $\beta_2=H_c/H_0>1$.
From equation (\ref{frweq4}), we see that the standard cosmology is recovered when $H_0<H(z)< H_c$ if $\beta_2\gg 1$.
At very early times when $H(z)\gg H_c$, the universe evolves according to $H^3 \sim \rho$. During radiation dominated era,
the universe evolves faster according to $a(t)\sim t^{3/4}$ instead of $t^{1/2}$.

If the parameters $\alpha_3$ and $\alpha_4$ take the particular values in equation (\ref{1parm}),
then Friedmann equation (\ref{frweq4}) becomes
\begin{equation}
\label{frweq5}
\begin{split}
3\beta_1\beta_2^2(1+\alpha)^2E(z)+3\beta_2[\beta_2^2-\beta_1\alpha(1+\alpha)]E^2(z)
+\beta_1\alpha^2E^3(z)\\
=3\beta_2^3\Omega_{m}(1+z)^{3(1+w_m)}
+\beta_1\beta_2^3(3+3\alpha+\alpha^2),
\end{split}
\end{equation}
with
\begin{equation}
\label{equiveq}
\beta_1=\frac{3(1-\Omega_m)\beta_2^3}{(\alpha^2+3\alpha+3)\beta_2^3-3(\alpha+1)^2\beta_2^2+3\alpha(\alpha+1)\beta_2-\alpha^2}.
\end{equation}
The deceleration parameter is
\begin{equation}
\label{qzeq3}
q(z)=-1+\frac{3\beta_2^3\Omega_m(1+w_m)(1+z)^{3(1+w_m)}}{\beta_1[\alpha E(z)-(1+\alpha)\beta_2]^2E(z)+2\beta_2^3 E^{2}(z)}.
\end{equation}
When $\beta_2\gg 1$, for most values of $\alpha$, $\alpha^2\beta_1/\beta_2^3\sim (1-\Omega_m)/\beta_2^3$ is negligible, therefore
the cubic term $E^3(z)$ can be neglected, which means that the model is not sensitive to the parameter $\alpha$.
The same is true for the most general case, so we expect that the model parameters $\alpha_3$ and $\alpha_4$ are not well constrained by
the observational data at the background level.

\section{Observational constraints}

To find out the parameters which are consistent with observatioal dat, we use the SNLS3 SNe Ia \citep{snls3}, BAO \citep{wigglez} and WMAP7 data \cite{wmap7}. The
SNLS3 SNe Ia data consists of 123 low-redshift SNe Ia data with $z\lesssim 0.1$
mainly from Calan/Tololo, CfAI, CfAII, CfAIII and CSP,
242 SNe Ia over the redshift range $0.08<z<1.06$ observed from the SNLS \citep{snls3},
93 intermediate-redshift SNe Ia data with $0.06\lesssim z\lesssim 0.4$ observed during the first season of Sloan Digital Sky
Survey (SDSS)-II supernova (SN) survey \citep{sdss2}, and 14 high-redshift SNe Ia data with $z\gtrsim 0.8$
from Hubble Space Telescope \citep{hstdata}.
To fit the SNL3 data, we need two additional nuisance parameters $\alpha$ and $\beta$
in addition to the model parameters $\mathbf{p}$.
For the BAO data, we use the measurements of the distance ratio $d_z=r_s(z_d)/D_V(z)$ at the redshift $z=0.106$ from the 6dFGS \citep{6dfgs},
the measurements of $d_z$ at two redshifts $z=0.2$ and $z=0.35$ from the distribution of galaxies \citep{wjp},
and the measurements of the acoustic parameter $A(z)=D_V(z)\sqrt{\Omega_m H_0^2}/z$ at three redshifts $z=0.44$, $z=0.6$ and $z=0.73$ from the WiggleZ dark
energy survey \citep{wigglez}.
In the fitting of BAO data, we need to use two more nuisance parameters $\Omega_b h^2$ and $\Omega_m h^2$.
For the WMAP7 data \cite{wmap7}, we use the measurements of the three derived quantities: the shift parameter
$R(z^{*})$, the acoustic index $l_A(z^{*})$ and the recombination redshift $z^{*}$.
The nuisance parameters $\Omega_b h^2$ and $\Omega_m h^2$ are again needed when we employ the WMAP7 data.
We apply the Monte Carlo Markov Chain code \citep{cosmomc,gong08} to find out the best fitting parameters.

As we discussed above, for the case $\alpha_3+4\alpha_4=0$, when $\beta_2\gg 1$, the model is almost equivalent to $\Lambda$CDM model and the model
is not sensitive to the values of $\alpha_3$ and $\beta_2$, so the value of $\beta_2$ is not bounded from above. By fitting the model to the observational data, we find that the minimum value
of $\chi^2$ for $\beta_2<1$ is much bigger than that for $\beta_2>1$. Therefore, the parameter space with $\beta_2<1$ can be ignored and
we choose $\beta_2>1$. Because $\beta_2$ is unbounded from above, we cut $\beta_2$ at $e^{10}$.

\begin{figure}
\includegraphics[width=0.8\textwidth]{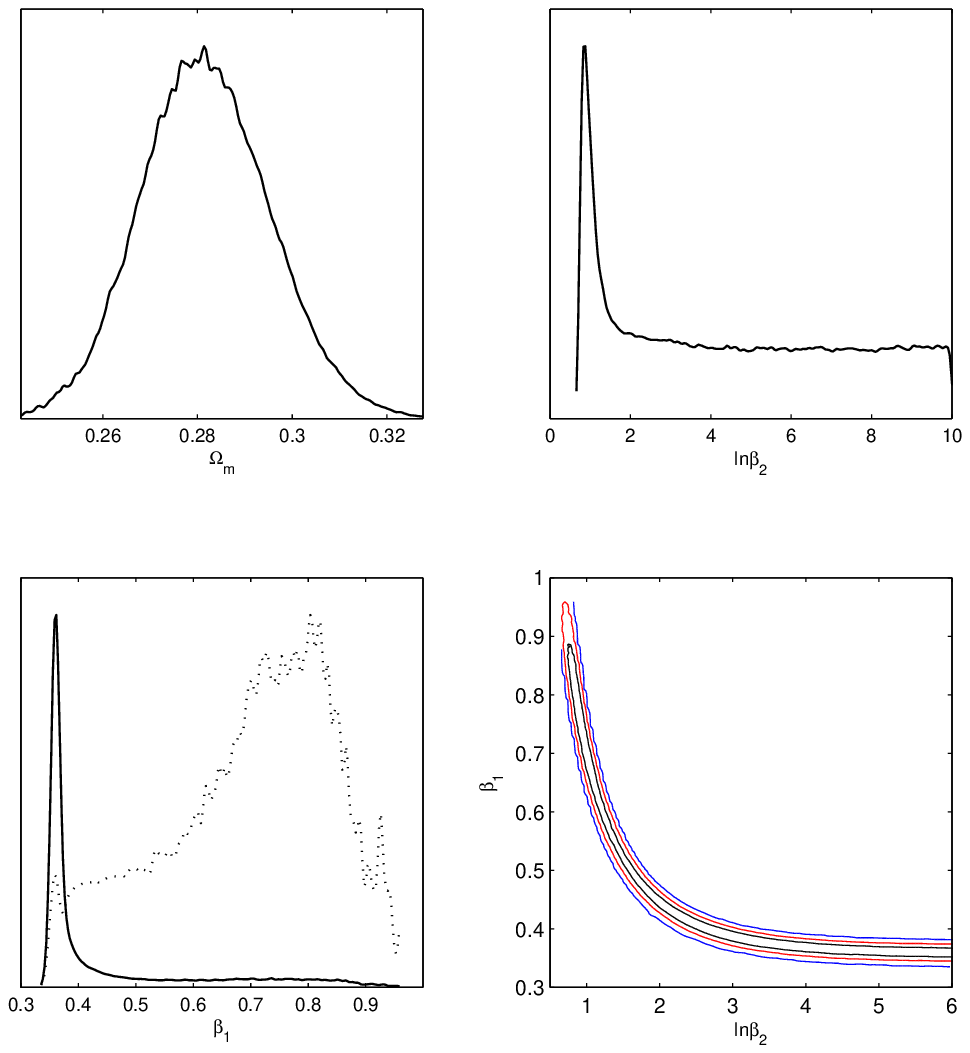}
\caption{The marginalized probability distributions of the model
parameters $\Omega_m$, $\ln \beta_2$ and $\beta_1$ and the marginalized contours of $\beta_1$ and $\ln\beta_2$.
The dotted line in the lower left panel shows the mean likelihood distribution.}
\label{b2likes}
\end{figure}

\begin{figure}
\includegraphics[width=0.8\textwidth]{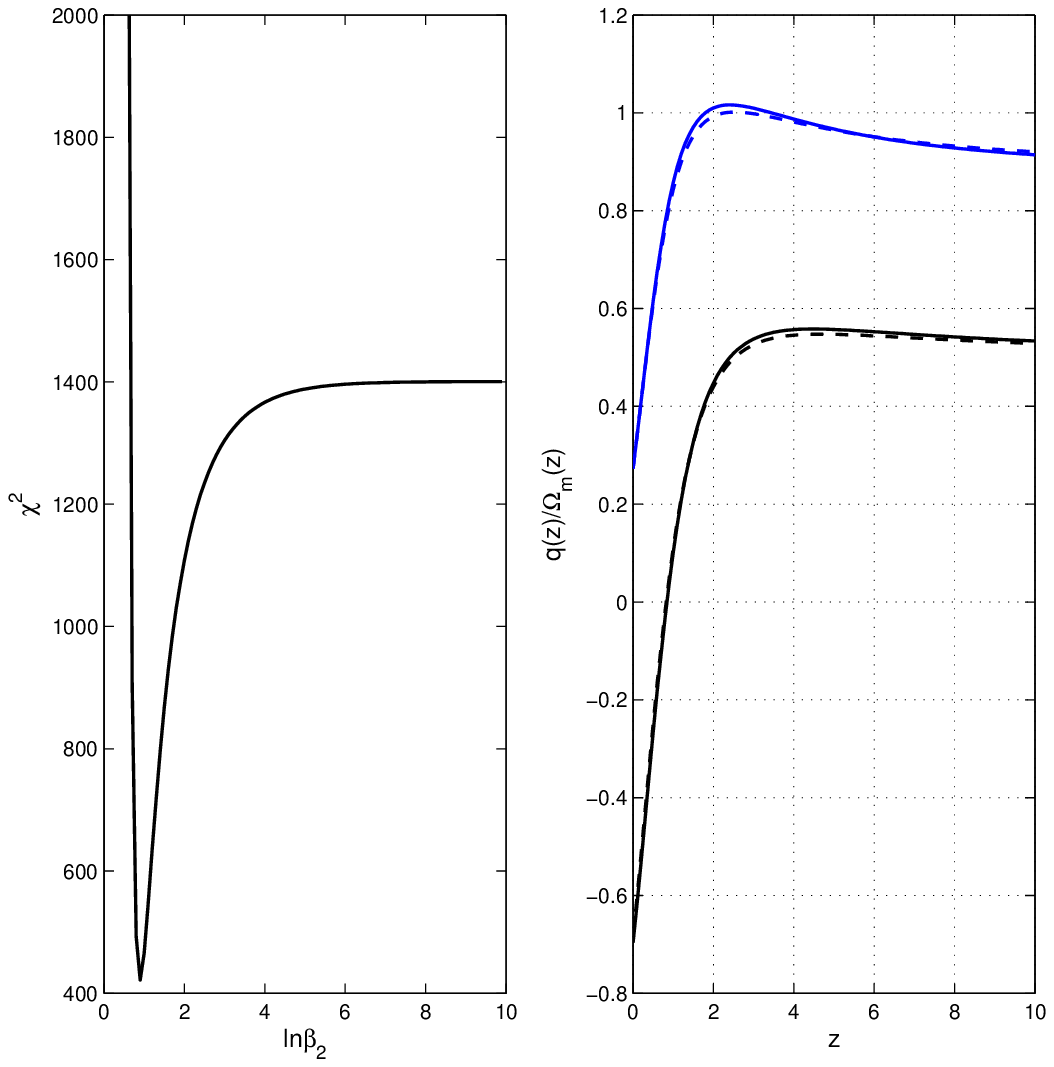}
\caption{Left panel: $\chi^2$ versus $\ln \beta_2$ for the special case $\alpha_3=\alpha_4$. The other parameters are fixed at their best fit values.
Right panel: The evolutions of $q(z)$ (black lines) and $\Omega_m(z)$ (blue lines), the solid lines are for the model with $\alpha_3=\alpha_4=0$
and the broken lines are the model with $\alpha_3+4\alpha_4=0$.}
\label{chisqb2}
\end{figure}

When $\alpha_3=\alpha_4=0$, fitting the model to the observational data, we get the minimum value of $\chi^2=421.65$, the best fit values $\Omega_m=0.27$ and $\beta_2=2.44$
and $\beta_2>1.93$ at $3\sigma$.
Note that for the curved $\Lambda$CDM model, the minimum value of $\chi^2$ is 423.98.
The marginalized probability distributions of the model
parameters $\Omega_m$, $\ln \beta_2$ and $\beta_1$ along with the marginalized contours of $\beta_1$ and $\ln\beta_2$ are shown in Fig. \ref{b2likes}. As expected, we get a long flat tail
for large $\beta_2$ and the most probable marginalized value of $\beta_1$ is around the asymptotical value $\beta_1\sim 0.37$, so the mass of graviton is around $0.6 H_0$.
From the mean likelihood distribution (the dotted line), we see that $\beta_1$ is peaked at its best fitted value. Because $\beta_2$ is not Gaussian, so the derived quantity
$\beta_1$ is not Gaussian either. On the other hand, $\beta_2$ is highly peaked at its best fit value. To see this point clearly, we plot the function of $\chi^2$ versus $\beta_2$
in Fig. \ref{chisqb2} by fixing the other parameters at their best fit values. From the plot in Fig. \ref{chisqb2}, we see that the probability of $\beta_2$ is negligible when
it is away from its best fit value. Note that this is different from the marginalized probability distribution because we neglect the degeneracies between $\beta_2$ and
other parameters including the nuisance parameters $\Omega_b h^2$ and $H_0$. Due to the two effects discussed above, the peak value of $\beta_1$ in the marginalized likelihood distribution
is different from that in the mean likelihood distribution. Using the best fit values, we plot the evolutions of the deceleration parameter $q(z)$ and the matter
density $\Omega_m(z)=8\pi G\rho_m/(3H^2)$ in Fig. \ref{chisqb2}.

\begin{figure}
\includegraphics[width=0.8\textwidth]{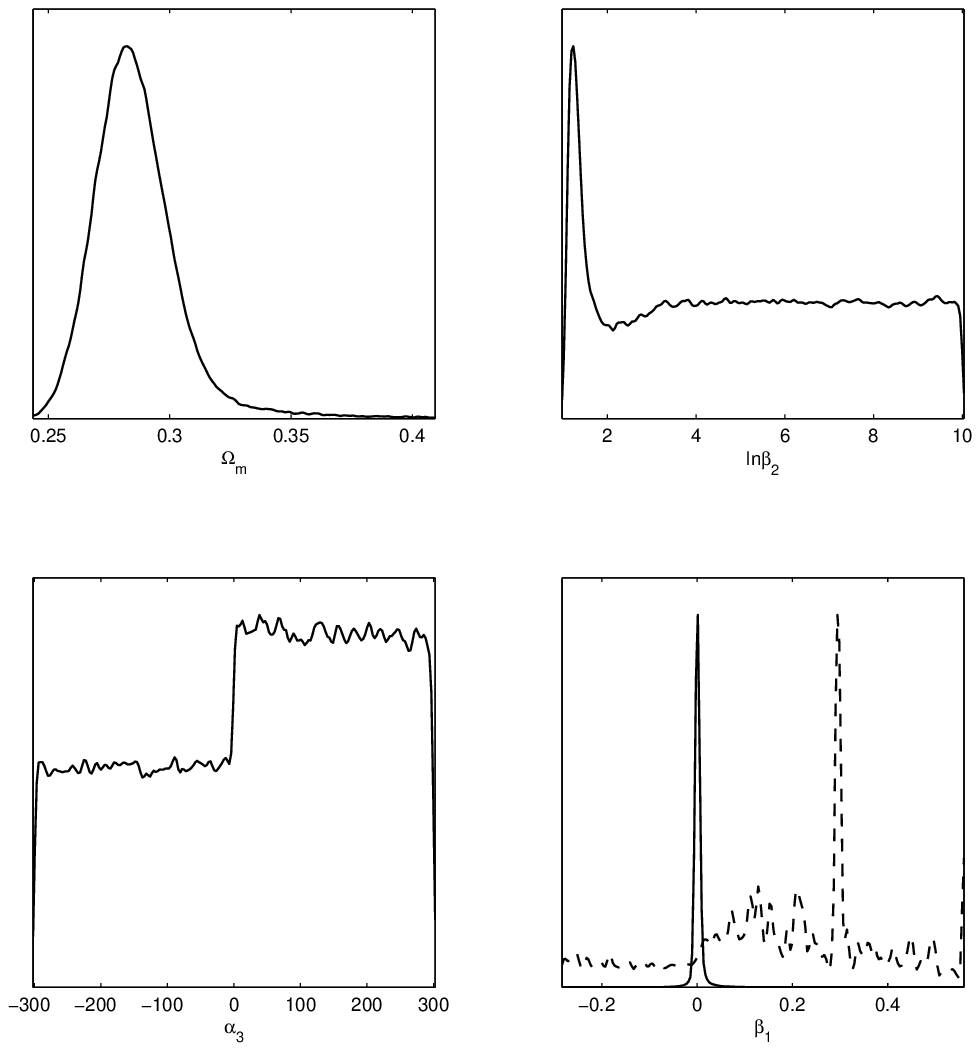}
\caption{The marginalized probability distributions of the model
parameters $\Omega_m$, $\ln \beta_2$, $\alpha_3$ and $\beta_1$. The dotted line in the lower right panel shows the mean likelihood distribution.}
\label{b2a3likes}
\end{figure}

For the case $\alpha_3+4\alpha_4=0$, we find that the best fit values are $\Omega_m=0.27$, $\beta_2=3.1$ and $\alpha_3=0.95$ with $\chi^2=421.57$.
The marginalized probability distributions of the model
parameters $\Omega_m$, $\ln \beta_2$, $\alpha_3$ and $\beta_1$  are shown in Fig. \ref{b2a3likes}. As expected, the model weakly depends on the value of $\alpha_3$, and nonnegative
values are slightly more probable than negative values.  From the marginalized probability distribution of $\beta_1$, we see that the most probable value is $\beta_1\sim 0$ which
means the graviton is almost massless. This can be understood from equation (\ref{b1eqcase2}), when $\beta_2\gg 1$ and $|\alpha_3|\gg 1$, $\beta_1\sim 0$.
For the same reason of non-Gaussian distribution, in the mean likelihood distribution (the dotted line), $\beta_1$ is peaked at its best fitted value.
With the best fit values, we construct the evolutions of the deceleration parameter $q(z)$ and the matter
density $\Omega_m(z)$ and they are shown in Fig. \ref{chisqb2}.

As discussed above, the parameters $\alpha_3$ and $\alpha_4$ are uncorrelated with $\Omega_m$ and $\beta_2$ for the general case, 
so we donot expect that the parameters $\alpha_3$ and $\alpha_4$ can be constrained by the observational data, and the results are similar to that of the special case $\alpha_3=\alpha_4=0$.
If the parameters $\alpha_3$ and $\alpha_4$ take the particular value (\ref{1parm}), the best fitting values are $\Omega_m=0.28$, $\ln \beta_2=14.42$ and $\alpha=-71.3$ with $\chi^2=424.28$.
Since $\beta_2\gg 1$, the probability distributions of $\beta_2$ and $\alpha$ are almost flat as expected. 
For the most general case, the best fitting values are $\Omega_m=0.28$, $\ln \beta_2=15.83$, $\alpha_3=-475.4$ and $\alpha_4=-362.1$ with $\chi^2=424.28$.
The fitting result for general $\alpha_3$ and $\alpha_4$ is almost the same as that when $\alpha_3$ and $\alpha_4$ take the particular values,
and $\alpha_3$ and $\alpha_4$ show a flat distribution, the plots are shown. Taking into
account the results obtained for the special case, we see that the special case with $\alpha_3+4\alpha_4=0$ fits the results better.

\section{Conclusions}

The cosmological constant solution (\ref{efflambda}) for different case was found by different group with different method by assuming the reference
metric to be Minkowski. Therefore the solution (\ref{efflambda}) should be true in the most general case with arbitrary spatial curvature and any model parameters.
In fact, the solution was found to be true in the most general case  for an isotropic and homogeneous universe by taking the ansatz (\ref{stuckphi1}) for the tensor $\Sigma_{\mu\nu}$.
Furthermore, new cosmological solution which modified Frriemann equation was also found in this approach. Therefore, the new solution
should also be a general solution and more cosmological solutions can be found following this approach \citep{gong12}.

Fitting the model to the observational data, we find that the best fit values $\Omega_m=0.27$ and $\beta_2=2.4$ with $\chi^2=421.65$ when $\alpha_3=\alpha_4=0$.
We also find that $\beta_2>1.93$ at $3\sigma$ level and the mass of graviton is around $0.6H_0$. For the special case $\alpha_3+4\alpha_4=0$,
we find that the best fit values $\Omega_m=0.27$, $\beta_2=3.1$ and $\alpha_3=0.95$ with $\chi^2=421.57$, which is almost the same as the case with $\alpha_3=0$.
In fact, we find that $\alpha_3$ is almost uncorrelated with the parameters $\Omega_m$ and $\beta_2$. For the most general case, $\alpha_3$ and $\alpha_4$ are
uncorrelated with the parameters $\Omega_m$ and $\beta_2$ and therefore are not constrained by current observational data at least at the background level.
The simple case with $\alpha_3+4\alpha_4=0$ is slightly favored by the observational data, so for phenomenological interest, we can consider the simple case only.

Although the model fits the observation as well as the standard $\Lambda$CDM model does, the phenomenology of the model is distinctively different from
dark energy models. As seen in Fig. \ref{chisqb2}, the matter density exceeds the critical density around redshift $z=2$ which may make the model testable,
and the period of this over-density is longer for the model with more degrees of freedom ($\alpha_3\neq 0$).
The consequence of the feature and how to detect massive gravity from astrophysical observations need to be further studied. 

\acknowledgments

This work was partially supported by
the National Basic Science Program (Project 973) of China under
grant No. 2010CB833004, the NNSF of China under grant Nos. 10935013 and 11175270,
CQ CMEC under grant No. KJTD201016, and the Fundamental Research Funds for the Central Universities.

\bibliography{massgravref} % all references

\end{document}